\documentstyle[prb,aps,epsfig]{revtex}
\begin{document}
\draft
\twocolumn[\hsize\textwidth\columnwidth\hsize\csname@twocolumnfalse%
\endcsname

\title{Microscopic Models of Two-Dimensional Magnets with Fractionalized
  Excitations}
\author{Chetan Nayak and Kirill Shtengel\cite{Irvine}} 
\address{Department of
  Physics and Astronomy, University of California at Los Angeles, Los
  Angeles, CA 90095-1547} 

\date{\today}
\maketitle
\vskip 0.5 cm

\begin{abstract}
  We demonstrate that spin-charge separation can occur in two
  dimensions and note its confluence with superconductivity, topology,
  gauge theory, and fault-tolerant quantum computation. We construct a
  microscopic Ising-like model and, at a special coupling constant
  value, find its exact ground state as well as neutral spin $1/2$
  (spinon), spinless charge $e$ (holon), and $\text{Z}_2$ vortex
  (vison) states and energies.  The fractionalized excitations reflect
  the topological order of the ground state which is evinced by its
  fourfold degeneracy on the torus -- a degeneracy which is unrelated
  to translational or rotational symmetry -- and is described by a
  $\text{Z}_2$ gauge theory.  A magnetic moment coexists with the
  topological order.  Our model is a member of a family of
  topologically-ordered models, one of which is integrable and
  realizes the toric quantum error correction code but does not
  conserve any component of the spin.  We relate our model to a dimer
  model which could be a spin $\text{SU}(2)$ symmetric realization of
  topological order and its concomitant quantum number
  fractionalization.
\end{abstract}

]
\narrowtext

\section{Introduction.}
The advent of polyacetylene \cite{Su80} and the fractional quantum
Hall effect \cite{Tsui82,Laughlin83} showed that quantum number
fractionalization is a robust possibility in condensed matter physics.
The quantum numbers of the low-energy excitations of these systems are
fractions of those of the microscopic degrees of freedom, the
electrons. There are charge $e$, spin $0$ and charge $0$, spin $1\over
2$ {\em spin-charge separated} excitations in polyacetylene and other
$1D$ systems.  The fractional quantum Hall state at filling fraction
$\nu=1/m$ has charge $e/m$, statistics $\pi/m$ excitations; more
exotic possibilities lurk at other $\nu$.  Despite a flurry of
interest generated by the suggestion \cite{Anderson87,Kivelson87} that
spin-charge separation is the mechanism for high-temperature
superconductivity in the cuprates, it is, at present, unclear whether
spin-charge separation can occur in a $2D$ magnet. There is a set of
long-wavelength field-theories
\cite{Balents98,Balents99,Balents00,Senthil99,Read91a,Read91b} which
describe the properties of putative fractionalized magnets, but their
existence has been controversial for want of a concrete microscopic
model of spin $1\over 2$ moments coupled by short-ranged interactions
in which fractionalization occurs. In this paper, we construct such a
microscopic model of a $2D$ magnet. We find the exact ground state and
neutral, spin $1\over 2$ (spinon) and charge $e$, spinless (holon)
excited eigenstates as well as a $\text{Z}_2$
vortex\cite{Kivelson89,Read89c,Wen91b,Senthil99}.  The fractionalized
excitations reflect the {\em topological order}
\cite{Wen90a,Wen90b,Wen91b} of the ground state which is evinced by
its fourfold degeneracy on the torus
\cite{Haldane88,Bonesteel89,Read89c} -- a degeneracy which is
unrelated to translational or rotational symmetry -- and is described
by a $\text{Z}_2$ gauge theory \cite{Fradkin79,Moessner99,Senthil99}.
Our construction implies that fractionalization is a reasonable
possibility for magnets with short-ranged interactions.  Our model is
a member of a family of models, another of which is integrable and
realizes the toric quantum error correction code \cite{Kitaev97}. The
models are related to the quantum dimer model \cite{Rokhsar88} and lie
at the confluence between superconductivity, topology, gauge theory,
and fault-tolerant quantum computation.

Our purpose here is to show that such microscopic models do exist, at
least in principle, so we construct a model with the aim that it be
deep within a phase supporting fractionalized excitations, not that it
be a realistic description of any particular physical system.
(As we discuss below, Kitaev \cite{Kitaev97},
in beautiful work, has constructed an exactly
soluble model with many of the desired properties, but it
does not have any conserved quantum numbers, so it is
not `fractionalized' in the sense of admitting
fractional quantum numbers.) 
However, we insist that our model be expressed in terms of spin-$1/2$
electrons, so that it is truly microscopic.  Consequently, our
analysis differs in a number of key respects from earlier ones which
dealt with models \cite{Read91a,Read91b,Balents98,Senthil99} which are
not, strictly speaking, microscopic electronic models or else relied
on various assumptions \cite{Rokhsar88,Balents00} in to order reduce
the microscopic models to effective models which exhibit
fractionalization.  We avoid the need for such assumptions or
modifications (however benign they may seem) by endowing our model
with the following properties which distinguish it from other models
which have been considered in this context: (1) Ising symmetry, (2)
translational symmetry which is broken by hand, and (3) adiabatic
continuability to an integrable model \cite{Kitaev97}.

A real magnet will have many additional complications,
but these are unimportant so long as it shares the key feature of our
model, namely {\it topological order}.  In pioneering work, Wen
\cite{Wen90a,Wen90b,Wen91b} observed that phases of matter in two
dimensions with fractionalized excitations are not characterized by a
local order parameter, in contrast to more familiar phases such as
crystals. Rather, their universal properties are encapsulated by
topological quantum numbers, such as their ground state degeneracy on
a torus or higher genus surface, {\it over and above any degeneracy
which is due to broken symmetry}. Degeneracy which is due to
topological order persists in the presence of local perturbations such
as impurities, which break translational and rotational symmetry.
(This observation will prove important since it guides us to construct
our model on lattices which penalize states which would break
translational and rotational symmetry on a square lattice.) This is
completely different from the two-fold degeneracy associated with an
Ising antiferromagnet, which is removed by the application of a small
symmetry-breaking field at even one point.  Topological order is
well-established theoretically in the fractional quantum Hall effect
\cite{Wen90b}, where it is manifested by the existence of excitations
with non-trivial braiding statistics \cite{Wilczek90}.

Along with spinons and holons, a spin-charge separated state must have
$\text{Z}_2$ vortices \cite{Kivelson89,Read89c,Wen91b}, which have
been recently dubbed `visons' \cite{Senthil99}.  Topological order
implies the existence of a gap in the vison spectrum.  It is {\it not}
necessary for all other excitations to be gapped (see, for instance,
the construction of Ref.~\onlinecite{Balents98}).  This is analogous
to the situation in a conventional ordered state such as a
superconductor, which can have gapless quasiparticles if, for
instance, it has $d$-wave symmetry or impurities.  They do not
preclude a stable SC state so long as there is a gap to the creation
of vortices.  Similarly, in our topologically-ordered states the
existence of a vison gap, $\Delta_v$, is necessary to guarantee the
existence of distinct topological sectors of the Hilbert space on the
torus, as we will describe below.  We compute the vison gap and
present evidence that the rest of the spectrum is gapped (though, we
reiterate, this is not a major issue). The integrable model in the
family is fully gapped.

The concept of topological order is very attractive theoretically
because it is precise, but it is sobering to note that it has not been
possible, to date, to directly measure most of the topological quantum
numbers -- such as the braiding statistics -- of a fractional quantum
Hall state.  On the other hand, this very feature has generated
considerable interest in the use of topologically-ordered states for
quantum computation. The inaccessibility of topological degrees of
freedom to local probes insulates them against many forms of
decoherence, the {\it b\^ete noir} of the quantum computation program.

This point was made by Kitaev \cite{Kitaev97} in a beautiful paper in
which he constructed a concrete model exhibiting the requisite
topological order and a fault-tolerant quantum error correcting code
which could be implemented in it (see also
Ref.~\onlinecite{Preskill97}). The integrable model in our family
is equivalent to Kitaev's. For our purposes, the model of greater
physical interest is the one which conserves $S^z$ and exhibits
quantum number fractionalization, which is of intrinsic interest and
might be relevant to high-temperature superconductivity
\cite{Anderson87}. It could also prove useful for quantum computing
since their spin and charge quantum numbers allows for the
manipulation of spinons and holons. Harnessing the otherwise elusive
visons also becomes a real possibility if the proposed experiment of
Ref.~\onlinecite{Senthil00a} can be implemented. Finally and perhaps
most importantly, the energy scale associated with topological order
in a magnet is likely to be an exchange constant $J\sim 1000K$. Thus,
a magnet with fractionalized excitations has many attractive features
as a milieu for quantum computation (for another, see
Ref.~\onlinecite{Freedman00} and refs.~therein).

\section{The Model.}  
Our model has spin $1\over 2$ degrees of freedom, ${{\bf S}_\alpha}$,
living on the links of a lattice which we specify below. They are {\it
  not} gauge fields, but gauge-invariant, physical degrees of freedom
which happen to be located on the links of the lattice (we return to
this point later). The Hamiltonian is:
\begin{equation}
  \label{eqn:f*}
  {H_0} = {J_1} {\sum_i}\,g\left({S^z_i}\right)
  \:\:  - \:\: {J_2} {\sum_p}{F_p}\,{P_p}\:\:
  + \:\: {J_3} {\sum_p}{P_p}
\end{equation}
where ${S^z_i} \equiv {\sum_{\alpha\in {\cal N}(i)}} {S^z_\alpha}\,$,
and ${\cal N}(i)$ is the set of links emanating from site $i$.  The
definitions of ${F_p}$, ${P_p}$ are:
\begin{eqnarray}
  {F_p} \equiv {\prod_{\alpha\in p}}{S^x_\alpha}
\end{eqnarray}
\begin{eqnarray}
  {P_p} \equiv
  f\!\left({S^z_{\alpha_1}}\!+\!{S^z_{\alpha_2}}\right)\cdot
  f\!\left({S^z_{\alpha_2}}\!+\!{S^z_{\alpha_3}}\right)\cdot
  f\!\left({S^z_{\alpha_3}}\!+\!{S^z_{\alpha_4}}\right)
\end{eqnarray}
where ${\alpha_1},{\alpha_2},{\alpha_3},{\alpha_4}$ are the links of
plaquette $p$, enumerated clockwise. At a site with coordination
number $z$, $g(x)={(2x+z-2)^2}/4$. $f(x)=1-{x^2}$. This model
is closely related to the quantum dimer model \cite{Rokhsar88}
(please see below).

\begin{figure}[htb]
  \begin{center}
    \includegraphics[width=1.6 in]{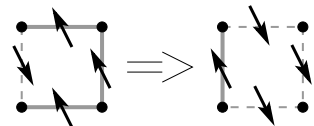}
    \caption{ The action of the flip
      operator $F_p$ on a typical plaquette. Notice that the total
      $z$-component of spin is generally not conserved under such
      operation. The links with the up-spins are shown here as colored
      -- this provides an alternative graphical representation which
      will be exploited later on.  }
    \label{fig:flip}
  \end{center}
\end{figure}

\begin{figure}[hbt]
  \begin{center}
    \includegraphics[width=1.6 in]{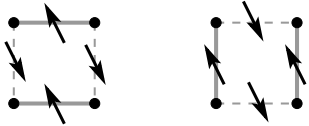}
    \caption{The action of the projection operator $P_p$ leaves
      plaquettes with the above shown spin configurations intact while
      annihilating {\em any} other type of plaquettes. A subsequent
      application of the flip operator $F_p$ to these plaquettes
      simply transforms them into each other, therefore they will be
      referred to as ``flippable''.  }
    \label{fig:projection}
  \end{center}
\end{figure}

The operator $g\left({S^z_i}\right)$ annihilates states (and only
those states) which have ${S^z_i}=-1$, i.e. which have one and only
one neighboring spin.  The operator $F_p$ `flips' plaquette $p$ by
flipping the four spins around it; an example of such flip is shown in
Fig.~\ref{fig:flip}.  The operator $P_p$ is a projection operator
which annihilates all states except those in which up- and down-spins
alternate around $p$ -- see Fig.~\ref{fig:projection}. Plaquettes are
assumed to have four sides.  however, they can be put together
irregularly or can overlap, as the parallelogram-shaped plaquettes of
the triangular lattice do.

We will take ${J_1},{J_2},{J_3}>0$.

\section{The Lattice.}  
Some care is required in the choice of lattice.  As we will see below,
the model (\ref{eqn:f*}) is tractable at ${J_2}={J_3}$. We would like
to choose a lattice so that the ground state exhibits the key feature
from which all of the interesting physics follows: fourfold ground
state degeneracy on the torus even in the presence of local
translational-symmetry-breaking fields.

This can be accomplished if (1) the lattice does not allow accidental
symmetries which will increase the ground state degeneracy, a
requirement which can usually be satisfied by taking a non-bipartite
lattice and (2) the lattice has a unit cell which includes several
plaquettes, so as to frustrate states in which the up-spins form an
ordered crystal (i.e.\ a spin-density-wave).  There are many possible
lattices which satisfy these requirements. Our basic strategy for
constructing these lattices is to take a Bravais lattice and introduce
a periodic array of `defects'. These defects pin a spin-density-wave
state and make it non-degenerate, but they do not affect the fourfold
degeneracy of the topologically-ordered state.  We arrange these
defects with a spacing which is incommensurate with the likely
spin-density-wave states so that these states are frustrated and
lifted in energy.  Certain types of defects will also make it easier
to satisfy (1).

\begin{figure}[hbt]
  \begin{center}
    \includegraphics[width=2.25 in]{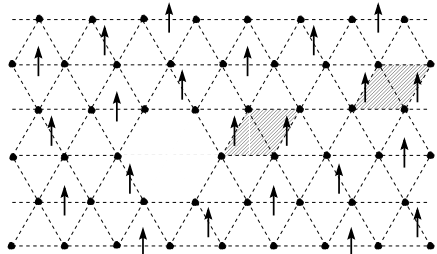}
    \caption{The mutilated triangular lattice, $T'$. 
      The spins (only up-spins are shown) correspond to the maximally
      staggered configuration. In the presence of this type of lattice
      defects, there are strings of flippable plaquettes (shaded) which
      frustrate the true staggered state.}
    \label{fig:latts}
  \end{center}
\end{figure}

\begin{figure}[hbt]
  \begin{center}
    \includegraphics[width=2.5 in]{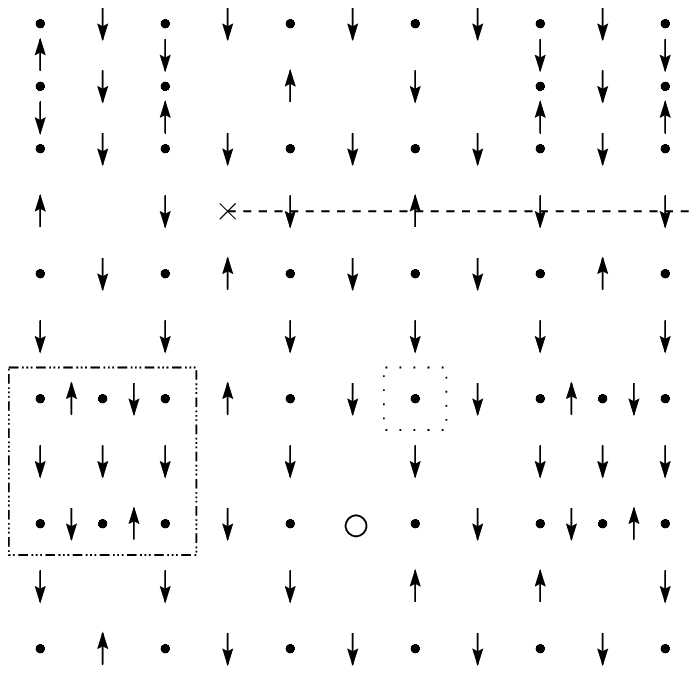}
    \caption{A state of our model on the distorted
      square lattice with a spinon (centered about the site
      inside the dotted line), a holon (open circle),
      and a vison (a cross connected to a dashed line).
      The dotted-dashed line encloses one of the
      four defect plaquettes which has been split by additional sites.}
    \label{fig:def_latt}
  \end{center}
\end{figure}

We will give two examples, $T'$ and $S'$.  $T'$ is based on the
triangular lattice (which, without defects, was exploited in this
context by Moessner and Sondhi \cite{Moessner00}, see below); the
defects are missing sites, as depicted in Fig.~\ref{fig:latts}.
Even a single such missing site frustrates the staggered state -- a
special type of crystalline state with no ``flippable'' plaquettes --
as depicted in Fig.~\ref{fig:latts}. (A ``flippable'' plaquette is a
plaquette which is not annihilated by $P_p$ -- see
Fig.~\ref{fig:projection}.  It has alternating up- and down-spins
whose direction can be reversed by application of ${F_p}{P_p} \equiv
S_{\alpha_1}^+ S_{\alpha_2}^- S_{\alpha_3}^+ S_{\alpha_4}^- +
\textit{h.c.}$. Left to its own devices, $F_p$ will flip any
plaquette, even the ones which are not `flippable'; $P_p$ prevents
this.  On the triangular lattice, a plaquette is any primitive
parallelogram.).  In $T'$, an array of sites is missing, so that the
lattice is given by
\begin{equation}
\Bigl\{ {\bf R} \Big|\: {\bf R}= {n_1} (a {\bf \hat x}) + {n_2}\left(
\frac{a}{2}{\bf \hat x} + \frac{\sqrt{3}a}{2}{\bf \hat y}\right)
\: ;\: {n_1},{n_2} \not\equiv 0\, \text{mod}\, k
\Bigr\}
\end{equation}
where $k$ is an arbitrary integer (this is just one such example, many
other $T'$-type lattices can be constructed along these lines).
Another possibility, $S'$, is the square lattice in which some of the
plaquettes are split in two, as in Fig.~\ref{fig:def_latt}
\cite{footnote_1}.  This must be done so as to split some plaquettes
horizontally and others vertically, in order to frustrate staggered
states aligned in both directions. These split plaquettes may be
viewed as elementary dislocations in a perfect square lattice and thus
they serve to model ``real-life'' defects.

\section{Ground States and Topological Order.}  
The first term in (\ref{eqn:f*}) requires each site to have one and
only one neighboring up-spin, so that ${S^z_i}= -1$, in which case it
vanishes.  On the square lattice, this leads to a magnetic moment
which is half that of a fully polarized state. In our model, this
magnetic moment does not result from spontaneous ordering;
the magnetization is actually fixed at a non-zero value.
However, it is possible to have a system which will
spontaneously undergo an Ising-like transition
into such phase \cite{Shtengel-prep}.  We will call
a state with both spin-charge
separation and ferromagnetism $F^*$, following the nomenclature of
Ref.~\onlinecite{Balents99}.  The coexistence of conventional
long-range order and quantum number fractionalization is familiar in
$1D$ and $2D$: in polyacetylene, fractionalization coexists with CDW
order; in easy-axis magnetic chains, with antiferromagnetism; at the
$\nu=1/3$ quantum Hall plateau, charge $e/3$ quasiparticles can form a
crystal and the topological order (and quantum Hall effect) will not
be disrupted.

The ground state of this model on $T'$ or $S'$ can be found exactly
for ${J_2}={J_3}$.  Every plaquette costs zero energy, so long as all
flippable plaquettes are taken in the linear combination
$|\psi\rangle+{F_p}|\psi\rangle$ \cite{Rokhsar88}.  Since every spin
configuration is obtainable from every other one by the repeated
application of ${F_p}P_p$ \cite{footnote_1,footnote_2}, the ground
state is the superposition with equal amplitudes of all possible
configurations of spins satisfying ${S^z_i}=-1$. The ground state is
annihilated by $H_0$.

Let us consider the crystalline states which compete with the
topologically-ordered state.  On the square and triangular lattices
\cite{Moessner00}, the staggered state state does not mix with other
states under ${F_p}P_p$ since it is annihilated by this operator.
It is a zero-energy ground state which is degenerate with
the topologically ordered state.
These two distinct ground states, have become
degenerate at the first-order phase transition point ${J_2}={J_3}$.

Fortunately, this is not the case on our lattices, as we
now demonstrate. The staggered
state has finite energy density at ${J_2}={J_3}$ on $T'$ and $S'$
since it is frustrated on these lattices.
On $T'$, there is no perfect staggered state.
Consider a single defect (missing site). It is clearly
impossible to have a perfect staggered state in the presence
of the defect. The closest that we can come to a perfect
staggered state (a which we will call a maximally staggered state)
is either (a) a state which has a string of
flippable plaquettes originating at the defect and extending
to infinity or (b) a state with one vertex which is frustrated
by having ${S^z_i}=-2$. If we take $J_1 \rightarrow \infty$,
then only (a) is possible.
When we introduce an array of defects, the strings of
flippable plaquettes will
originate at one defect and terminate at another.
Under the action of $F_p$, this state will mix with all
of the others. Hence, a maximally staggered state will
have energy density proportional to $\sqrt{\rho}\,{J_2}$ in this limit
(where $\rho$ is the defect density).
However, for $J_2 \rightarrow \infty$, only (b) is possible
and the energy density of the maximally staggered state is
proportional to $\rho\,J_1$. More generally, for $J_1$ and $J_2$
finite, the energy density of the maximally
staggered state will be $\rho{J_1}/2$ for small defect density $\rho$
and will be proportional to $\sqrt{\rho}\,{J_2}$ at large $\rho$.
One might wonder whether there is some other crystalline
state (e.g. one with a large unit cell)
which has zero energy. However, if such a state
contains `flippable' plaquettes, it will mix under
the action of $F_p$ with all of the other states with
flippable plaquettes \cite{footnote_2}. Hence, such a state
will cost finite energy.

We can repeat the above analysis for $S'$.
On $S'$, we can frustrate one plaquette for each defect, with energy
density $\rho J_2$.  On the perfect square lattice, the $F^*$ ground
state is not fourfold degenerate. It is critical \cite{Rokhsar88}, and
unstable to a columnar state as $J_3$ is decreased \cite{Sachdev89}.
This is not the case on $T'$ or $S'$, so we do not need to worry about
the columnar state, either.

The ground state may be visualized in the following way. Consider some
reference configuration of spins which is annihilated by the first
term of equation~(\ref{eqn:f*}) and color all of the links which have
up-spins.  Now take any other configuration which is also annihilated
by the first term of equation~(\ref{eqn:f*}) and do the same. By
placing one graph on top of the other, and erasing all links at which
both graphs coincide, we obtain a collection of loops on the lattice.
If we visualize states in terms of their associated loop graphs, then
the ground state is given by a superposition of different loop
configurations.

Since $H_0$ conserves modulo $2$ the winding numbers of these loops
about either of the generators of the torus, there are $4$ degenerate
ground states on the torus, ${\psi_{\left({n_1},{n_2}\right)}}$,
${n_1},{n_2}=0,1$, with $0,1$ corresponding to even or odd winding
numbers.  By straightforward extension, the degeneracy on a genus $g$
surface is $4^g$. Although we have computed this degeneracy only at
the special coupling constant value ${J_2}={J_3}$, we believe that it
is robust over some range of parameters because it is characterized by
an integer, $4$. This integer cannot change as a result of
infinitesimal perturbations, but only as a result of a perturbation
which is sufficiently strong that it moves the system across a phase
transition at which this integer changes discontinuously.

On the perfect square lattice, the directed winding number
(not merely the winding number modulo $2$) is conserved
because the lattice is bipartite. As a result,
there are $L\times L$ sectors. $S'$ is not bipartite,
so it has only 4 topologically-distinct ground
states.\cite{footnote_3}

\section{Spinons and Holons.}  
The fourfold ground state degeneracy implies the existence of
fractionalized excitations, so long as it is unrelated to
translational and rotational symmetry, a condition which is satisfied
as a result of our choice of lattice. To see this, imagine cutting
open the torus along its second generator, thereby producing an
annulus. The ground state ${\psi_{\left(0,0\right)}}$ has non-zero
projection on the ground state of the annulus because it will have
some amplitude to have zero loops encircling the torus.
${\psi_{\left(1,0\right)}}$ does not, but it does have finite
projection on a finite-energy excited state of the annulus because it
must have at least one loop circuiting the torus and it must have some
amplitude to have only one.  This finite-energy excitation has, by
construction, a spinon at the inner and outer edges of the annulus, as
in Laughlin's construction of charge $e/3$ quasiparticles in the
fractional quantum Hall effect \cite{Laughlin83}.  The inner edge of
the annulus can be shrunk and filled in since our discussion depends
only on the topology, but not the geometry of the lattice.  This
construction can be done on a torus of any size, so the spinon at the
boundary can be taken arbitrarily far away from the one in the
interior with finite energy cost.

This general argument can be substantiated in our model by a direct
construction of spinon and holon excitations.  Two spinons may be
created by flipping a single up-spin into a down-spin. This changes
${S^z}$ by $-1$ and creates $2$ sites with ${S^z_i}=-2$.  These sites
can be moved apart; each one carries ${S^z}=-{1\over 2}$ and costs
energy ${J_1}$.  A holon may be constructed by simply removing a spin
from one of these spinon sites.  This removes charge $e$ and spin
${S^z}=-{1\over 2}$ from a neutral ${S^z}=-{1\over 2}$ excitation,
thereby producing a spinless charge $e$ excitation. There will now be
a site with ${S^z_i}=-{3\over 2}$, so the holon costs energy
${J_1}/4$.  In the loop picture, spinons and holons reside at the
endpoints of broken loops.

\section{Visons.}
Consider now the operator
\begin{equation}
  \label{eqn:vison_def}
  {\Phi_p} \equiv {\prod_{\alpha\in {c_p}}}2{S^z_\alpha}
\end{equation}
where $c_p$ is any curve which starts at the center of plaquette $p$,
connects it to the center of a neighboring plaquette $p'$, and
continues in this manner through the centers of a sequence of
neighboring plaquettes, running to infinity (or the boundary of the
system). The product in equation~(\ref{eqn:vison_def}) is over all
links $\alpha$ which intersect $c_p$. Under the action of $\Phi_p$,
each loop configuration receives a $-1$ if $c_p$ has an odd number of
intersections with colored links and $1$ if it has an even number of
intersections with colored links.  When a holon or spinon follows a
trajectory encircling $p$, the intersection number must change by one,
so ${\Phi_p}$ creates a $\text{Z}_2$ vortex, or `vison'
\cite{Senthil99}, at plaquette $p$.

The statistics of holons and spinons depend on the energetics of the
model: by binding to a vison, they can switch their statistics between
bosonic and fermionic \cite{Kivelson89,Read89c}.
Our Hamiltonian does not allow holons or
spinons to move: they are infinitely heavy. However, a small
perturbation will allow them to move and will give rise to the
energetics which determines whether they bind with visons and, thereby,
their statistics.

The fourfold degeneracy of the ground state -- or, in other words, the
topological order -- guarantees the existence of a vison energy gap.
To see this, consider the degenerate states ${\psi_{(0,0)}} \pm
{\psi_{(1,0)}}$ on the torus.  Now imagine creating a vison pair at
plaquette $p$, taking one vison around the second generator of the
torus, and annihilating the pair at $p$. This is equivalent to acting
on our state with an operator similar to $\Phi_p$, but with the curve
$c_p$ in equation~(\ref{eqn:vison_def}) replaced by a closed curve
which passes through $p$ and encircles the torus along its second
generator. This operator exchanges ${\psi_{(0,0)}} \pm
{\psi_{(1,0)}}$. The amplitude for such a process is essentially the
exponential of the Euclidean action required for such a virtual
process to occur, $\sim {e^{-c L \Delta_v}}$ where $L$ is the length
of the loop around the torus, $\Delta_v$ is the vison gap, and $c$ is
a constant.  Hence, the energy splitting between states
${\psi_{(0,0)}}$ and ${\psi_{(1,0)}}$ is $\sim {e^{-c L \Delta_v}}$.
Since we know that this splitting vanishes in the thermodynamic limit,
the vison gap $\Delta_v$ must be finite.

This conclusion is supported by a direct calculation.  The creation of
a vison at $p$ takes the state $|\psi\rangle+{F_p}|\psi\rangle$ into
$|\psi\rangle-{F_p}|\psi\rangle$, with an energy cost $\Delta_v$.
Since the vison creation operator $\Phi_p$ commutes with all of the
terms in equation~(\ref{eqn:f*}) except for the $J_2$ term at
plaquette $p$, with which it anticommutes, a state with one vison,
$\left|{\Phi_p}\right\rangle$, has excitation energy:
\begin{equation}
  \label{eqn:vison_E}
\left\langle{\Phi_p}\right|\: H\:\left|{\Phi_p}\right\rangle =
  2{J_2}\: \langle 0|\: {P_p}\:|0\rangle
\end{equation}
Hence, the vison gap is equal to $2{J_2}$ multiplied by the density of
flippable plaquettes. This may be computed at ${J_2}={J_3}$ by the
Grassmann techniques discussed below. In an integrable model which we
discuss below, exact vison eigenstates and energy eigenvalues may be
found.

{}From equation~(\ref{eqn:vison_E}), we see that the vison gap will be
non-vanishing whenever $\langle 0|{F_p}{P_p}|0\rangle\neq 0$, i.e.\ 
whenever the spins fluctuate in the ground state, as they generically
do in our model, even outside the topologically ordered phase.
However, this is not particularly consequential.  Consider the
analogous situation in a superfluid: it is possible to define a vortex
energy above the transition (e.g.\ the Kosterlitz-Thouless transition)
which varies smoothly across the transition.  However, this energy is
only meaningful in the superfluid state (or, perhaps, near it).
Similarly, the vison gap becomes meaningful in the topologically
ordered phase. Outside this phase, $\Phi_p$ is merely an operator
which creates some complicated gapped excitation.

A vison gap is necessary for topological order; a gap in the rest of
the spectrum is not.  However, the equal-time spin-spin correlation
function in the ground state is exponentially decaying, as may be seen
from an exact mapping between the ground state of our model and the
field theory of free lattice fermions \cite{Samuel80a}, according to
which it is a square root of the eight-fermion correlator. This decays
exponentially with distance since these fermions, though massless on a
regular square lattice, acquire mass in the presence of lattice
distortions such as that shown in Fig.~\ref{fig:def_latt} (the details
of this calculation will be published elsewhere).  Hence, it is
natural to conclude that there is an energy gap to all excited states;
this may be argued via the single-mode approximation.  This
computation has been carried out on the triangular lattice by Moessner
and Sondhi \cite{Moessner00}, who found that the single-mode
approximation suggests that the system is indeed gapped.

\section{A Family of Models.}  
We can gain further insight into the spectrum of our model by
generalizing it to the following family of models:
\begin{eqnarray}
  \label{eqn:h'}
  H = {H_0} &+& {J'_1} {\sum_i}\,y\left({S^z_i}\right)
  \nonumber \\
  { } &-& {J'_2} {\sum_p}\left[{F_p}\,\left(1-{P_p}\right)
    - \left(1-{P_p}\right)\right]
\end{eqnarray}
where $y(x)=\frac{2}{3} x\left({x^2}-4\right)(x+1)$ (for $z=4$) and
${J_2}={J_3}$ in $H_0$.  As ${J'_1},{J'_2}$ are increased, two things
occur: loop crossings are allowed \cite{footnote_4} and plaquette
flips of `unflippable' plaquettes are allowed.
The topological order
is preserved because the winding number modulo $2$ is conserved. In
the extreme limit ${J'_1}={J_1}$, ${J'_2}={J_2}$, there are eight
equally likely configurations at each site (corresponding to those of
the eight-vertex model) and the model is now integrable because
$[g\left({S^z_i}\right)+y\left({S^z_i}\right)]$ and $F_p$ commute
among themselves and, hence, with the Hamiltonian, for all $i$, $p$.
Hence, we can simultaneously diagonalize all of these operators. They
have eigenvalues $0$ and $1$, respectively, in the ground state; a
$g\left({S^z_i}\right)+y\left({S^z_i}\right)=1$ eigenvalue is a
quasiparticle excitation at $i$ and ${F_p}=-1$ is a vison at $p$.
This model is equivalent to Kitaev's \cite{Kitaev97}.  The ground
state of this integrable model has the same topological order
(fourfold degeneracy) as that of (\ref{eqn:f*}), but its quasiparticle
excitations do not carry spin since it is not conserved. Since there
is no projection operator $P_p$ in $H$, the vison energy is exactly $2
J_2$.  Crystalline states have energy ${J_2}/2$ per plaquette above
the ground state.

\section{Field Theoretic Description.}  
The configurations allowed in the ground state of the integrable model
are described by closed loops -- in other words, by the configurations
of the Ising model on the dual lattice (equivalent to the eight-vertex
model). The dynamics of the plaquette flip operator is the same as
that of a transverse field in the Ising model. Hence, the integrable
model is equivalent at low energies to the transverse field Ising
model which, in turn, is dual to a $\text{Z}_2$ gauge theory.  Since
the topological order associated with $H_0$ is the same as that of the
integrable model, it, too, is described by a $\text{Z}_2$ gauge theory
\cite{Fradkin79}, as proposed in
refs~\onlinecite{Moessner99,Senthil99}.

Note that $S^z_i$ is conserved for all $i$ in our model
(\ref{eqn:f*}). Hence, there is an independent $\text{U}(1)$ symmetry
at each site of the lattice. However, only time-independent
transformations leave the Lagrangian invariant.  Hence, this is an
ordinary (but large) symmetry group; it is {\it not} a gauge symmetry,
which must allow time-dependent transformations.  All of the degrees
of freedom in (\ref{eqn:f*}) are physical. This is similar to the
symmetry of a set of noninteracting spins in a magnetic field,
$H=-{\sum_i} {\bf B}\cdot {\bf S}_i$, which also has an independent
$\text{U}(1)$ at each site of the lattice.

\section{Quantum Dimer Models.} 
Our model (\ref{eqn:f*}) can be mapped to the quantum dimer model
\cite{Kivelson87,Rokhsar88,Kivelson89,Fradkin90}, in which it is
assumed that that there are spins located at the sites of a lattice
and that each spin forms a singlet dimer with one of its nearest
neighbors. In (\ref{eqn:f*}), an up-spin link corresponds to a dimer;
a down-spin link to the absence of a dimer; spinons, to empty sites
(which are holons in the dimer model).  Then the first term in
equation~(\ref{eqn:f*}) requires each spin to form a dimer with
exactly one of its neighbors.  The $J_2$ and $J_3$ terms are precisely
the dimer kinetic and potential energies of
Ref.~\onlinecite{Rokhsar88}.  Our $F^*$ state of (\ref{eqn:f*}) is
simply the resonating valence bond (RVB) \cite{Anderson87,Rokhsar88}
ground state of the quantum dimer model on the same lattice.
Recently, Moessner and Sondhi \cite{Moessner00} gave compelling
evidence that the triangular lattice quantum dimer model has an RVB
ground state over a substantial range of parameters terminating at a
first-order phase transition at ${J_2}={J_3}$ into the staggered
state.  According to our arguments, the RVB state is the unique, exact
ground state at ${J_2}={J_3}$ on $T'$.

\section{$\text{SU}(2)$ symmetric models.}
There is no reason to believe that an $\text{SU}(2)$ symmetric magnet
cannot be topologically-ordered. If we wish to apply the preceding
results, then the quantum dimer model can be the starting point for
the discussion of an $\text{SU}(2)$ symmetric topologically-ordered
magnet \cite{Chayes89,Klein82,private}.  There is some numerical
evidence \cite{Misguich99} that a model of spins on the triangular
lattice with strong four-spin exchanges -- reminiscent of the
plaquette flip operator -- has a topologically-ordered ground state
with fourfold degeneracy.

\section{Summary.}  
We have demonstrated that quantum number fractionalization and
topological order are an eminently reasonable possibility for
two-dimensional magnets by constructing a microscopic model which
exhibits these phenomena.  Our result links two problems of great
interest: quantum number fractionalization in $2D$ quantum magnets and
the investigation of physical systems which are suitable platforms for
fault-tolerant quantum computation (this link was also noted in
Ref.~\onlinecite{Bonesteel00}).  A possible nexus with ideas about
superconductivity -- either in the cuprates or elsewhere -- leads to
potentially fruitful avenues for further research in both areas.

\begin{acknowledgments}
  We would like to express our gratitude to S.~Chakravarty, L.~Chayes,
  M.~Freedman, A.~Kitaev, R.~B.~Laughlin, R.~Shankar, R.~Rajaraman,
  T.~Senthil, and, especially, S.~Kivelson for valuable conversations.
  C.N. and K.S. are both supported by the National Science Foundation
  under Grant No.~DMR-9983544.  C.N.  is also supported by the
  A.~P.~Sloan Foundation.
\end{acknowledgments}


\begin{references}
\bibitem{footnote_1} {Notice that the introduction of certain types of
    lattice defects, such as those depicted in
    Fig.~\ref{fig:def_latt}, generates a few plaquettes (namely two
    for every split) that contain an odd number of spins.  These
    plaquettes should be considered ``unflippable'' since flipping
    them would break $S_z$ conservation.  Nevertheless every spin
    belonging to such plaquettes is also a member of another,
    flippable plaquette and therefore a low density of such defects
    will not result in ``frozen'' spins so long as some extra
    constraints are added.}
\bibitem{footnote_2} {This can be proven inductively by assuming that
    this is true on all $2,4,\ldots,2N$ site lattices with free
    boundary conditions if every site has at least two neighbors.  If
    we add a pair of sites satisfying the same condition, then we can
    show by direct construction that the resulting $2N+2$ site
    lattices will also have the property that all spin configurations
    are reachable by repeated application of ${F_p}{P_p}$.  This
    inductive step only fails when we add a pair of sites, each of
    which is part of the same two plaquettes and no others. This
    necessarily occurs if we try to impose periodic boundary
    conditions, and it signals the existence of winding number
    sectors.}
\bibitem{footnote_4} {As ${J'_2}$ increases, some ${S^z_i}=1$
    configurations will occur in the ground state.  Let us color
    up-spin links as before and superimpose the resulting picture on
    that of a reference state which only allows ${S^z_i}=-1$.  The
    ground state is again a sum over all loops, but with crossings
    allowed.}
\end{references}
\end{document}